# Science Requirements for EAGLE for the E-ELT

C. J. Evans[1], M. D. Lehnert[2], J.-G. Cuby[3], S. L. Morris[4], A. M. Swinbank[4], W. D. Taylor[1],
D. M. Alexander[4], N. P. F. Lorente[1], Y. Clénet[5] & T. Paumard[5]

[1] UK Astronomy Technology Centre, Royal Observatory Edinburgh, Blackford Hill, Edinburgh, EH9 3HJ, UK
[2] GEPI, Observatoire de Paris, 5 Place Jules Janssen, 92195 Meudon Cedex, France
[3] Laboratoire d'Astrophysique de Marseille, BP8, 13376 Marseille Cedex 12, France
[4] Department of Physics, Durham University, South Road, Durham, DH1 3LE, UK
[5] LESIA, Observatoire de Paris, 5 Place Jules Janssen, 92195 Meudon Cedex, France

**ABSTRACT**

We present an overview of the EAGLE science case, which spans spatially-resolved spectroscopy of targets from five key science areas – ranging from studies of heavily-obscured Galactic star clusters, right out to the first galaxies at the highest redshifts. Here we summarise the requirements adopted for the study and also evaluate the availability of natural guide stars in example fields, which will impact on the adaptive optics performance and architecture.

## 1. INTRODUCTION

EAGLE is a French-UK Phase A study of a multi-IFU, near-IR spectrometer for the E-ELT [1]. Its science case comprises spatially-resolved spectroscopy in five broad areas of astrophysical research:

- The physics and evolution of high-redshift galaxies;
- Detection and characterisation of first-light galaxies;
- The physics of galaxy evolution from stellar archaeology;
- The stellar content, mass functions and dynamics of stellar clusters;
- Co-ordinated growth of black holes and galaxies.

Each of these topics features in the list of nine 'prominent science cases' assembled by the E-ELT Science Working Group, now being used for detailed science simulations as part of the Design Reference Mission (DRM). Following a discussion of some general requirements, we introduce each of the science topics, highlighting specific requirements that arise from them.

### 1.1. General Requirements

In the Phase A study we have assumed that EAGLE will be mounted at the planned E-ELT Gravity Invariant Nasmyth Focus (GIFS), in which the instrument focal-plane is parallel to the ground (compared to a traditional on-axis Nasmyth focus, in which an instrument experiences a variable gravity vector as it rotates to compensate for field rotation). The unvignetted focal plane at the GIFS is expected to be a circular field with a diameter of ~5' (i.e. an area of approx. 20 arcmin$^2$), although as part of the Phase A study we are looking at methods to increase this slightly. To address the science topics listed above there are a number of general requirements:

- Spatially-resolved spectroscopy of multiple targets simultaneously, suggesting a system that uses integral-field unit (IFU) spectrographs combined with adaptive optics (AO) correction in the vicinity of each target.
- The size of each IFU should be well matched to the size of potential targets (typically in the range of 1-2"), which should also include a number of 'spare pixels' to enable local sky subtraction.
- Wavelength coverage from 0.8 to 2.45μm. This spans the primary diagnostic lines in high-redshift systems and also includes the calcium triplet for studies of stellar populations.

- A minimum spectral resolution of $R\sim4{,}000$, so that OH sky-emission lines can be well resolved and subtracted.
- The target acquisition system should be able to map contiguous areas, in addition to providing observations of clustered and relatively evenly distributed targets.

We now discuss the main components of the science case.

## 2. THE PHYSICS AND EVOLUTION OF HIGH-REDSHIFT GALAXIES

The assembly and evolution of the baryonic component of high-redshift galaxies does not simply follow the hierarchical merging of dark matter structures via gravitational collapse – other forces such as cooling, loss/exchange of angular momentum, and feedback due to star formation and active galactic nuclei (AGN) are important factors. Semi-empirical models have been developed to describe the formation and evolution of galaxies, but these are heavily reliant on simple parameterisations of the physical processes at work and are determined from relatively small observational samples. The critical ingredients in the models are metallicity, angular momentum, the stellar initial mass function (IMF) and the spatial distribution of the gas, as well as including the effects of galaxy interactions/mergers and the feedback rates from supernova explosions and super-massive black holes.

The best approach for testing these models is to measure spatially-resolved properties such as the star-formation histories, metallicities, extinction, clustering and dynamics of a wide range of individual sources and any satellite/companion galaxies. Instrumentation on 8-m class telescopes such as VLT-SINFONI has given us remarkable views of galaxies at excellent spatial resolution, e.g. [2], [3], and Figure 1. However the targets currently within our grasp are intrinsically biased toward the brightest/largest galaxies – only with the sensitivity of the E-ELT can we access a wider range of targets to gain a clear understanding of the high-redshift galaxy population. Even then, observational samples must be drawn from sufficiently large parent populations (numbering thousands) and over large enough volumes to avoid field-to-field variations ("cosmic variance") and temporal variations that may bias the results.

### 2.1. Source Densities

Science cases often quote source densities of galaxies from a specific selection technique. However, taking such densities at face value can be very misleading because the number of targets for a given science goal does not always overlap, which can lead to deceptively large numbers of potential targets. In reality, if one is interested in observations of multiple spectral lines to estimate gas-phase abundances, e.g. [O II] $\lambda\lambda 3726$-29, [O III] $\lambda 5007$ and H$\beta$ (as well as H$\alpha$ to estimate the extinction), the number of suitable targets is much lower than those from published source densities.

Here we focus on the source density of objects in the redshift range $1.5<z<3.5$, over which most of the important rest-frame optical diagnostic lines fall within one of the near-IR bands. The number of objects per square arcminute ranges from one to five in various redshift ranges, as shown in Table 1. Therefore, in the context of EAGLE, a field-of-view of ~20 arcmin$^2$ would suggest a multiplex in the range of 20 to 100. However, as already discussed, the number of targets suitable for detailed investigation with EAGLE is not the full source density, but something smaller. Other populations at these redshifts such as QSOs, radio galaxies and sub-millimetre galaxies have even lower source densities than these optically and near-IR selected samples. To strike a good balance between scientific return and cost, ~20 deployable IFUs over a region of 20 arcmin$^2$ would satisfy many science goals for studies of galaxy evolution (with a corresponding increase in the multiplex if a larger patrol field can be obtained).

**Table 1: Source densities for galaxies in different redshift ranges**

| Object Type | Selection | Redshift Range | Surface Density | Ref. |
|---|---|---|---|---|
| BX galaxies | UGR colour, R<25.5 | $1.9 < z < 2.7$ | ~5-6 arcmin$^{-2}$ | [4, 5] |
| BM galaxies | UGR colour, R<25.5 | $1.5 < z < 2.0$ | ~3.1 arcmin$^{-2}$ | [4] |
| LBGs | UGR colour, R<25.5 | $2.7 < z < 3.4$ | ~1.7 arcmin$^{-2}$ | [5] |
| DRGs | J-K colour, K<22.8 | $2 < z < 3.5$ | ~1.6 arcmin$^{-2}$ | [6] |
| K-selected | K<23 | $2 < z < 2.5$ | ~1.2 arcmin$^{-2}$ | [7] |
| K-selected | K<23 | $2.7 < z < 3.3$ | ~1.3 arcmin$^{-2}$ | [7] |

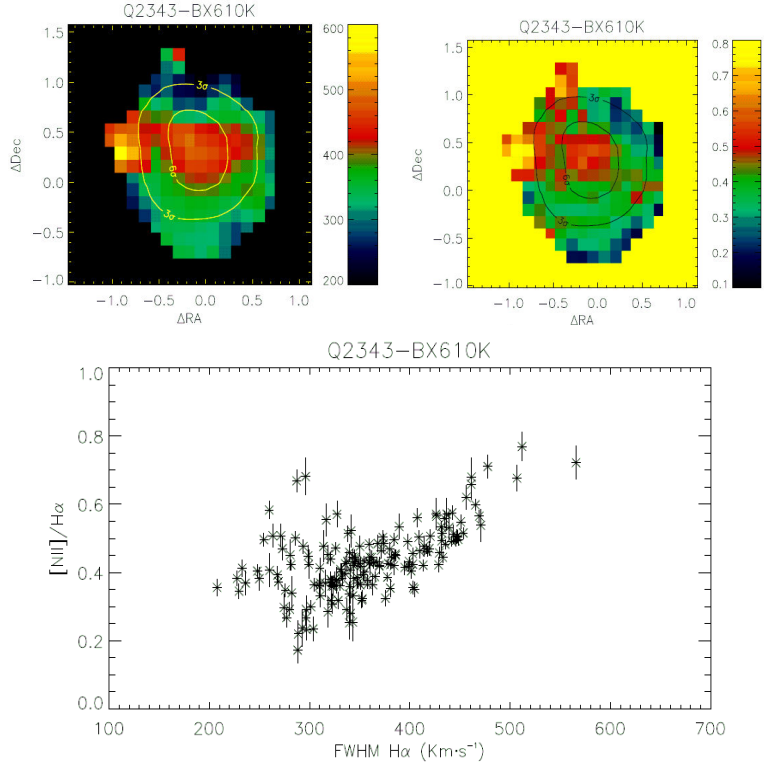

**Figure 1:** VLT-SINFONI results for a galaxy at z~2 [2]. The spatial correlation between line-width and line-ratio suggests strong mechanical heating of the nuclear and minor-axis gas, as seen in local starbursts with energetic superwinds. EAGLE will resolve such winds with greater spatial resolution to investigate their properties and estimate the amount of mass, metals and energy that such winds eject at high redshift.

### 2.2. Spatial Resolution

To investigate the spatial resolution required to recover meaningful results for high-redshift systems, we turn to simulations that employ Hα observations of nearby disk galaxies combined with results of simulated mergers [8, 9]. These were instigated using parameters relevant to the EAGLE study, and have since been undertaken as part of the effort toward the DRM. Example simulated images are shown in Figure 2. To probe the physics of galaxy formation *in situ* requires ≥30% ensquared energies (EE) in the H-band, over a sampling scale of ~100-150 mas. However, we are currently pushing this requirement towards smaller scales of 75 mas to probe the 'fine scale' structures within galaxies (e.g. Jean's radii, clumpy structures, bars, disk instabilities, etc.), as opposed to the more diffuse and low-surface-brightness emission. New AO simulations, combined with further science modelling, will be used to revisit this requirement in the next phase of the study.

### 2.3. IFU Field-of-view

The field seen by each IFU should be ≥1.5"x1.5", providing sufficient spatial coverage for most applications. For example, this would just encompass most of the galaxy shown in Figure 1 (for comparison, note that the simulated images in Figure 2 are 0.8" x 0.8"). Observations of larger targets would be improved by having larger sub-fields (~2", or more, on a side), particularly in the sense of having local background pixels included within each IFU.

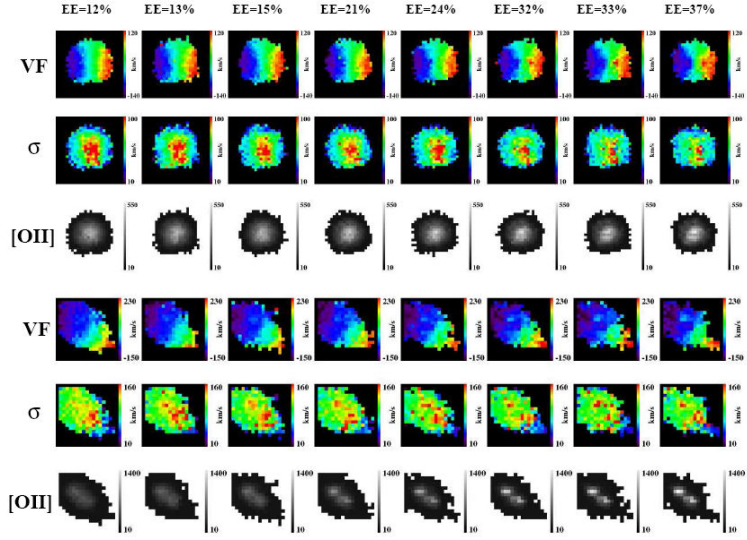

**Figure 2: Simulation of rotating disk (upper 3 panels) and merger (lower 3 panels) at z=4 with 50 mas pixels [8]. The panels show the recovered velocity field (VF), dispersion (σ) and the [O II] emission-line map for both scenarios. The ensquared energy (EE) increases from left to right, from 12 to 37% EE per 100 mas.**

## 3. DETECTION AND CHARACTERISATION OF FIRST-LIGHT GALAXIES

Understanding the basic properties of the first galaxies (7<z<~20) is one of the most important challenges in modern astronomy. In the standard cosmological model, dark matter halos arise from the gravitational collapse of primordial perturbations in the initial dark matter density distribution. Cooling is very efficient on these scales and gas quickly collapses to the centre of the halo, where it cools further to form the first stars and galaxies. These galaxies should have sufficiently large potential wells to retain photoionisation-heated and supernova-shocked gas, and therefore are able to self-regulate the process of star-formation for more than one generation of stars. Studies of the formation of these first stars, assembly of the first galaxies, and the growth of super-massive black holes through gas accretion is the key to obtaining a complete picture of star formation and quasar activity in the early Universe. This will help to identify the galaxies responsible for reionisation – were they rare but UV-intense AGNs, or less massive, more ubiquitous star-forming galaxies?

Observational constraints on the properties of galaxies at z~6 and beyond are scarce, with only a handful of galaxies confirmed via spectroscopy, e.g. [10]. Indeed, due to the faintness of these galaxies the observations are severely photon-starved and so current work focuses on confirming redshifts through identification of a single emission line (predominantly Ly-α 1215Å). As such, very little is known about their stellar and dynamical masses, or their stellar populations. With the huge collecting area of the E-ELT, one of the primary extra-galactic science goals is to probe the properties of the first galaxies in comparable detail to that achieved currently for lower-redshift galaxies.

### 3.1. Source Densities

We have used the VLT-SINFONI Exposure Time Calculator (v3.2.2) to provide an order-of-magnitude estimate of the flux detection limit achievable with the E-ELT. In 100 hours, SINFONI expects to reach a 10σ detection of the Ly-α λ1215 emission line (point source) with a line flux of $1\times10^{-18}$ erg/s/cm$^2$ (R=4,000, with 25 mas pixels, an intrinsic line width of ~50 km/s and assuming good AO correction). Scaling this by the difference in primary apertures (background-limited case) suggests a limiting flux of ~$2\times10^{-19}$ erg/s/cm$^2$.

Theoretical predictions of Ly-α emitters [11, 12] estimate that between z = 7.5 and 9.5 (500 Mpc/h), the density of galaxies with fluxes >5×10$^{-19}$ erg/s/cm$^2$ is ~10±5 per arcmin$^2$ (depending on cosmic variance). Of course, the number of EAGLE targets will also depend on the target selection technique, most likely from narrow-band imaging surveys. For a 1% filter centred at z=8.5, the space density then drops to less than 1 per arcmin$^2$. However, targets could be selected from pre-imaging that uses several narrow-band filters (i.e. at different redshifts), thereby increasing the number of targets. Broad-band drop-out techniques in the deepest survey fields will also boost the number of targets. The relatively low (expected) source densities mean that the notional patrol field of ~20 arcmin$^2$ is required as a minimum, preferably larger. When coupled with densities of 1 to 2 per arcmin$^2$ this suggests a multiplex of at least 20 channels.

### 3.2. Spatial Resolution

Although these galaxies are very small, diffraction-limited performance is not required. Spatial resolution of ~0.1"-0.2" scales will suffice, with 10 to 20 angular resolution elements across the target to attempt to probe the stellar populations within the galaxy. Moreover, many of the Lyman-break galaxies detected at z = 5-6 show multiple components or companions. If this trend continues to higher redshift (as expected in a hierarchical mass-assembly framework), then resolving the velocity offsets and stellar populations between multiple components/merging systems provides a route to constraining the dynamical masses and stellar mass-to-light ratios of these galaxies.

### 3.3. Clustering/Tiling Requirements

Searches for first-light galaxies could be undertaken as follow-up to deep narrow-band imaging, as a blind spectroscopic survey or, in large galaxy clusters which act as powerful gravitational lenses on the more distant systems lying behind them. The lensed case provides a good example of the configuration requirements for the EAGLE IFUs: distributed, clustered, and contiguous mapping. For example, IFUs can be configured to target all of the giant arcs in a cluster in a single exposure, exploring lensed systems at z = 1-5 (left-hand panel of Figure 3). At the same time, the central part of the field can be mapped to target the caustic region in search for first light (z = 7–20) galaxies, with other IFUs used to observe other faint arclets to accurately define the cluster-lensing potential (right-hand panel of Figure 3).

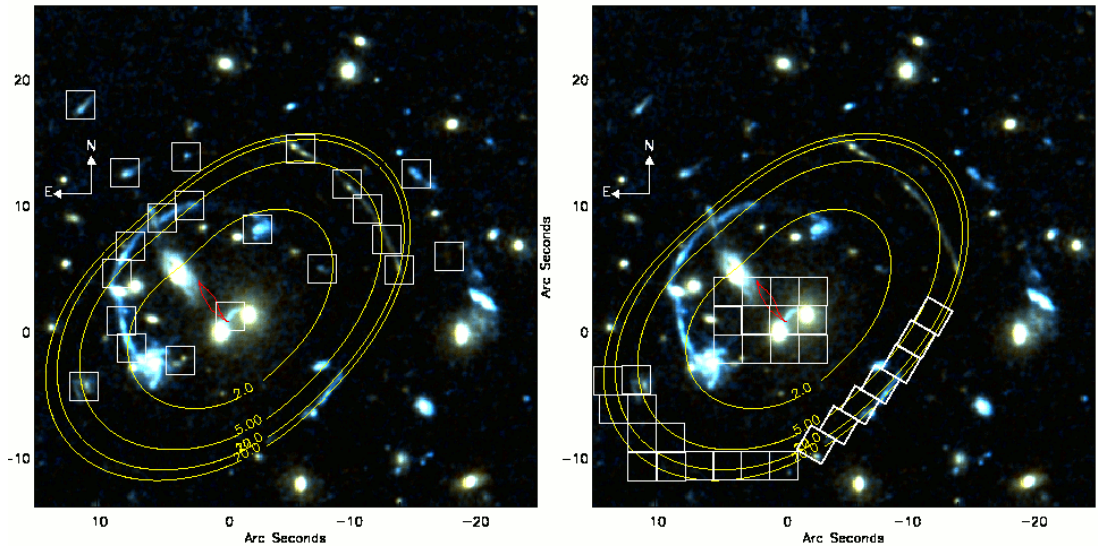

**Figure 3:** *HST* **image of the RCS0224-002 galaxy cluster at z = 0.78 with 2.2"x2.2" IFU footprints overlaid. Both panels show the high-redshift critical curves for z = 2, 5, 10 and 20. Left: Example EAGLE configuration to observe the giant arcs and arclets at z ~ 1 to 5. Right: Using EAGLE to map the critical curves, searching for highly magnified, very high-redshift galaxies.**

# 4. THE PHYSICS OF GALAXY EVOLUTION FROM STELLAR ARCHAEOLOGY

With its large primary aperture and excellent spatial resolution, the E-ELT will be *the* facility that enables us, for the first time, to undertake stellar spectroscopy in a wide range of galaxies in the Local Volume. This will open-up a much larger number of galaxies for study and, more importantly, will also encompass a much wider range of galaxy morphologies and metallicities compared to those in the Local Group. There is increasing evidence for the accretion of numerous low-mass satellite galaxies in the assembly of the present-day Milky Way, e.g. [13]; do we see evidence for similar processes at work in other large spiral galaxies? Moreover, what are the assembly and star-formation histories in galaxies with other morphological types, such as massive ellipticals and metal-poor irregulars? As we move beyond the Local Group, potential targets include:

- NGC 3109 and Sextans A with sub-SMC metallicities (both at the edge of the Local Group at 1.3 Mpc);
- Numerous spiral-dominated galaxy groups out to ~10 Mpc (e.g. Sculptor and M83 Groups);
- Centaurus A (NGC 5128, at 3.8 Mpc), the nearest elliptical;
- M82 (at 4.5 Mpc) which displays significant recent star formation and evidence for a galactic superwind;
- NGC 3379, the nearest 'normal' elliptical (at 10.8 Mpc);
- The Virgo Cluster of galaxies at ~15-17 Mpc, the nearest massive cluster.

Only by studying such a diverse range of systems can we begin to understand the evolution of the different populations of which galaxies are composed. Photometric observations are an immensely powerful method for studies of stellar populations in external galaxies, but only via measurements of chemical abundances and stellar kinematics can we definitively disentangle the different populations – i.e. follow-up spectroscopy is required.

## 4.1. Spectral Resolution

A higher-resolution mode (i.e. $R > 4,000$) is strongly desirable for quantitative stellar spectroscopy with EAGLE. Over the past decade the calcium triplet (CaT) has become a ubiquitous method to obtain estimates of metallicities and stellar kinematics, e.g. [14, 15]. The CaT comprises well-characterised lines arising from singly-ionized calcium, with rest wavelengths of $\lambda\lambda 850$, 854 and 866 nm. This adds a strong requirement for spectral coverage bluewards of 1 μm. Recent results from VLT-FLAMES have demonstrated that, with careful calibration, [Fe/H] estimates obtained from the CaT with low-resolution spectroscopy ($R$~6,500) are in agreement with direct measurements from higher-resolution spectroscopy ($R$~20,000) with the same instrument [16].

Selection of a higher-resolution mode is a trade-off between sensitivity and the precision with which one can recover robust abundance estimates. This trade-off is also strongly informed by the desire to obtain precise radial velocities. In preparation for the ESA-GAIA mission, there has been significant effort toward understanding the velocity precision in the CaT region as a function of spectral resolution, not to mention the computation and compilation of libraries of model atmospheres and empirical spectral templates. In particular, the effect of spectral resolution and signal-to-noise in CaT spectroscopy has been thoroughly investigated [17], with the results employed to quantify the uncertainties on velocities from the Radial Velocity Experiment (RAVE, [18]):

$$\log (\Delta RV) = 0.6*(log\ SNR)^2 - 2.4*log\ SNR - 1.75 log\ R + 9.36$$

where SNR is the signal-to-noise ratio of the observed CaT spectrum. This relation demonstrates the power of increasing $R$ compared to greater signal-to-noise, in terms of the velocity precision achieved. The typical SNR of the RAVE spectra is 30, at $R = 7,500$, so the predicted error is 2.2 km/s (although the actual performance is slightly better than this, given spectral sampling with more than 2 pixels per element and wider wavelength coverage than GAIA [18]). In Table 2 we summarise the spectral resolution, as a function of signal-to-noise ratio (SNR), required to obtain a velocity precision of 2 km/s.

Given that many of the E-ELT targets will be at the very limit in terms of their faint magnitudes, SNR > 20 is relatively optimistic. Adopting a more pragmatic SNR of ~15 gives the requirement for $R$=10,900. For simplicity, we have adopted $R$~10,000 as the baseline requirement for a high spectral-resolution mode for EAGLE. Note that the requirements for such velocity precision are, if anything, more strongly driven by observations in the Galactic Centre and of stellar clusters, where the expected velocity dispersions will be smaller.

**Table 2: The effects of velocity precision (ΔRV) and signal-to-noise ratio (SNR) on spectral resolution (*R*).**

| ΔRV [km/s] | SNR | *R* |
|---|---|---|
| 2 | 30 | 8,000 |
| 2 | 20 | 9,400 |
| 2 | 15 | 10,900 |
| 2 | 10 | 14,000 |

### 4.2. IFU Field-of-view

Although the primary objective is spectroscopy of individual stars, extended spatial-coverage via IFUs for each channel is strongly desirable at $R\sim10,000$. For instance, with fibre-based instruments such as VLT-FLAMES there is no chance of recovering spectral information if objects are partially blended. Moreover, IFUs of ~1.5" x 1.5" (thereby including spatial pixels outside of the seeing-disk), will have the benefit of local background subtraction – one of the problems that can limit the usefulness of emission-line spectroscopy with e.g. VLT-FORS in studies of extra-galactic stellar populations. Also, in some instances, multiple stars may be located within the same IFU, thereby enhancing the size of the observed sample

### 4.3. Spatial Resolution

In terms of both crowding and sensitivity/observing efficiency, we want the best possible spatial performance of telescope and instrument, combined with a large multiplex. For observations of the evolved red-giant populations in galaxies beyond the Local Group, seeing-limited or GLAO observations will not deliver sufficient contrast, particularly as the GLAO-correction in the region of the CaT will be less effective than at longer wavelengths. Given the challenges of AO correction in this wavelength region, a degree of pragmatism is required in specifying the spatial resolution. For now we adopt 30% EE in 200 mas – quantitative modeling of this specification is now underway (e.g. the cluster simulations in the next section).

## 5. THE STELLAR CONTENT, MASS FUNCTIONS AND DYNAMICS OF STELLAR CLUSTERS

The predominant factor determining the evolution of a star is its initial mass. The combination of state-of-the-art evolutionary models with empirical IMFs that describe the stellar mass distributions is a powerful tool to study coeval stellar populations; be they in small knots of star formation, in an open cluster, in a globular cluster, or across an entire galaxy. With the improved spatial resolution from the E-ELT we will be able to explore embedded clusters and star-forming regions in unprecedented detail throughout the galaxies of the Local Group. Moreover, the E-ELT will provide us with a view of the different components of massive star clusters beyond the Local Group for the first time, e.g. in M82, the nearest starburst galaxy. An integral part of such work is the feedback of observational results to population-synthesis models – by rigorously testing these in the local Universe, we can draw more meaningful conclusions when they are applied to observations of high-redshift, star-forming galaxies.

IFU spectroscopy will be invaluable to characterise the spectral content and velocity dispersions within massive clusters in the Local Group, e.g. in heavily-obscured Galactic clusters. This method is already being used with VLT-FLAMES to map dense Galactic clusters such as NGC 3603 (e.g. Figure 4); the E-ELT will be able to penetrate these more deeply.

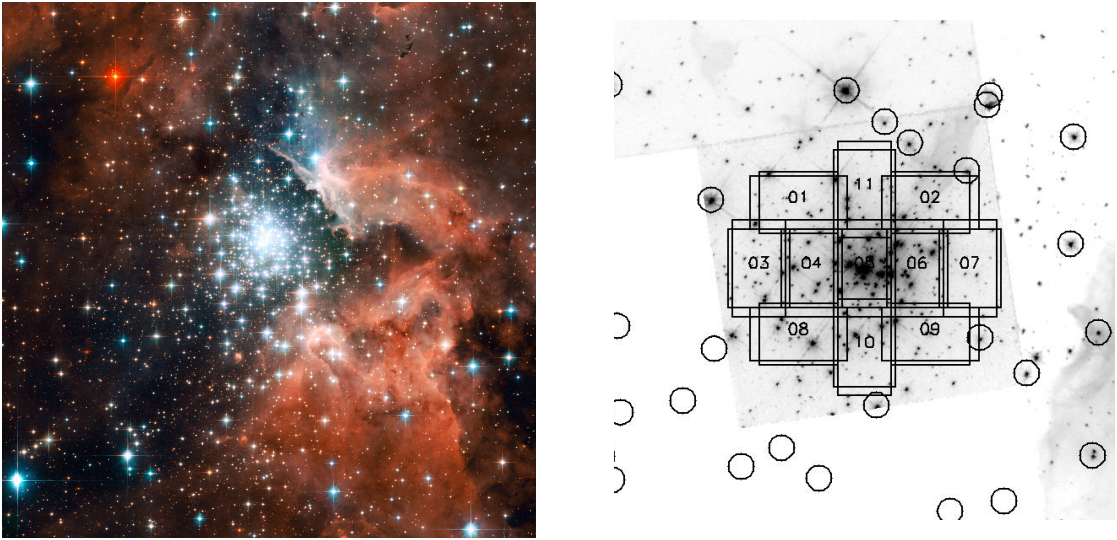

**Figure 4:** Looking for dynamical evidence of mass segregation in the Galactic cluster NGC 3603. Left: *HST*-ACS composite of the cluster; Right: The core region with FLAMES IFU positions overlaid (Gieles et al., in prep).

### 5.1. Spatial Resolution

In regions with low extinction, spectroscopy in the *I*-band will be the most sensitive region observed with the E-ELT; the background is low compared to longer wavelengths, and yet a useful correction can still be achieved from AO, thereby improving the sensitivity significantly. To investigate the I-band performance of potential AO systems on the E-ELT, we have begun simulations using NGC 346, the most massive stellar cluster in the Small Magellanic Cloud as a template. As a preliminary investigation, we have scaled the population of NGC 346 to the distance of NGC 3109, a metal-poor irregular galaxy at the edge of the Local Group at 1.3 Mpc. In this context we simply use the photometry and astrometry of the high-mass, luminous population [19], rather than employing much deeper *HST* imaging that is also available (which would be shifted several magnitudes beyond the detection limit of even the E-ELT).

Simulated images were generated using the Specsim IDL application [20]. Originally written to model imaging and spectroscopy with the *JWST* Mid-InfraRed Instrument (MIRI), Specsim has now been modified for use in development of the E-ELT and EAGLE science cases. At this stage we have used the software to (effectively) convolve an input target list with different AO PSFs to generate simulated I-band images, as shown in Figure 5. The images show the effects of crowding in a 1.6"x1.6" sub-field near the centre of NGC 346. The ground-layer and laser tomography AO PSFs (GLAO and LTAO; left- and right-hand panels of Figure 5) are from Miska LeLouarn's end-to-end simulations at ESO, while the multi-object AO PSF (MOAO) is from analytic calculations undertaken at ONERA as part of the EAGLE Phase A study. Each PSF assumes a Paranal-like turbulent profile, otherwise the main assumptions are:

- GLAO: 5 LGS at a radius of 3', with 0.8" seeing @ 0.5 µm;
- MOAO: 6 LGS at a radius of 3.4', with 0.95" seeing;
- LTAO: 5 LGS at a radius of 45" (with an additional LGS at the centre of the field), with 0.8" seeing.

It is clear that the GLAO mode (envisaged as the default operational mode of the E-ELT) is not sufficient to disentangle the components of such a population. Superficially, the LTAO performance appears better than that from MOAO (with the caveat that different models have been used), but the important issue is that the individual stars are sufficiently resolved by MOAO to characterise the population. This is coupled with the huge advantage that LTAO is limited to one pointing, whereas the MOAO architecture enables multiple observations over a large (>2', cf. MCAO) area.

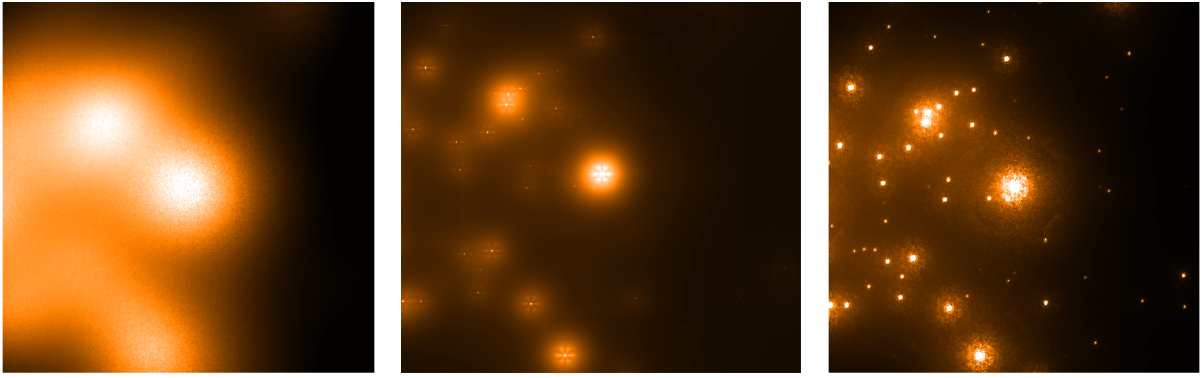

Figure 5: I-band simulated images (1.6" x 1.6") of a part of NGC 346, projected to a distance of 1.3 Mpc. Left: Simulated image using GLAO PSF; Centre: MOAO; Right: LTAO.

## 6. CO-ORDINATED GROWTH OF BLACK HOLES AND GALAXIES

Once considered rare and exotic phenomena, AGNs are now thought to play a crucial role in the formation and evolution of galaxies, e.g. [21, 22]. Key observational evidence for this is the finding that every nearby massive galaxy harbours a central super-massive black hole (SMBH; typically $> 10^6$ $M_{Sun}$) with a mass directly proportional to that of its spheroid, e.g. [23, 24], referred to as the $M_{BH}$–$M_{SPH}$ relationship. This seminal discovery indicates that all massive galaxies have hosted AGN activity at some time over the past ~13 Gyrs, and suggests that galaxies and their SMBHs grew concordantly, despite nine orders of magnitude difference in size scale.

Exploring how SMBHs grow, and the mechanism that regulates the $M_{BH}$–$M_{SPH}$ relationship across a wide range of environments, is a fundamental goal of cosmology. The most likely mechanism is via accretion-related winds, jets, and outflows, which provide an opportunity for the SMBH to orchestrate star formation in the host galaxy through feedback effects (heat/radiation pressure; shocks/mechanical energy). Unfortunately, we have limited observational evidence to constrain these processes in the distant Universe, when these effects were probably the most important. To complicate matters further, it is not clear whether the $M_{BH}$–$M_{SPH}$ relationship is constant throughout cosmic time, with suggestions that the ratio is higher in the high-redshift Universe than that found locally [25, 26]. However, all of these studies have focused on objects known to host massive SMBHs and may be strongly biased towards this result [27, 28]. It is therefore important to constrain directly the SMBHs and host-galaxy properties of distant typical AGNS, and to search for the presence of accretion-related outflows, across a range of environments.

There are two distinct EAGLE cases in the context of AGN science. The first is to observe high-z AGNs in survey fields to (1) map the spatial extent of the emission-line gas kinematics, constraining potential emission-line outflows and the dynamics of the host galaxy and nearby companions, and (2) to provide spectroscopic redshifts for the faintest (most distant) AGNs below the sensitivity limits (~5x10$^{-18}$ erg/s/cm$^2$) of current spectroscopic surveys, e.g. [29]. Without the increased sensitivity and improved spatial resolution of the E-ELT, little will be known about the emission-line properties of distant typical AGNs, with even less known about their emission-line dynamics. Studies of the distant AGN population with the E-ELT will provide insight into how SMBHs grow as a function of cosmic epoch and environment. For example, by studying AGN activity across a range of environments, from typical blank fields to distant protoclusters, it will be possible to determine whether accretion-related outflows, and interaction–major-merger events are more prevalent in overdense regions. A comparison of the properties of the AGN-hosting galaxies to non-AGN-hosting galaxies (cf. Sections 2 and 3) will also provide insight into what drives SMBH growth.

These science goals will require multi-deployable IFUs with sub-fields of ~2". For AGNs with large extended emission-line halos and in distant protocluster fields, it would be useful to place the IFUs close together to provide near contiguous mapping, thus allowing the emission-line gas kinematics to be traced over large linear scales. Given the wide distribution of source redshifts for AGNs discovered in blank-field X-ray surveys, it would also be advantageous to select a filter for each IFU to ensure that the same rest-frame line is studied for each object. With the large source densities of X-ray detected AGNs, it is beneficial to have one IFU per arcmin$^2$, although future multi-wavelength surveys might identify larger source densities so this is not a strong constraint.

The second EAGLE case is to observe local AGNs that have extended features such as star-formation rings and structures spiralling toward the nucleus, e.g. NGC 1097 (at 14.5 Mpc, Figure 6). Here we require relatively clustered IFU observations (or even contiguous mapping) of the core and its environs. Mapping the dynamics and stellar content in these structures will then improve our understanding of how the gas and dust inflows are channeled into the SMBH.

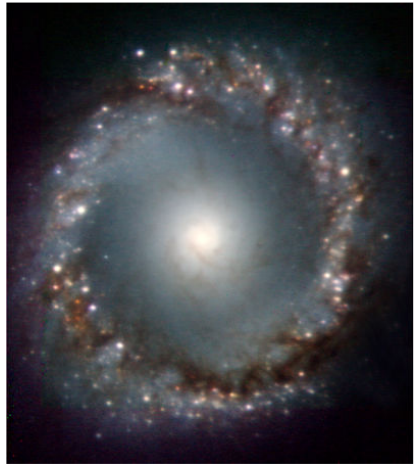

**Figure 6:** VLT NAOS-CONICA image of the active galaxy NGC 1097 (combined J, H and Ks, [30]). The field-of-view shown is ~25"x25", corresponding to a physical size of 1.7 x 1.7 kpc.

## 7. AVAILABILITY OF NATURAL GUIDE STARS

To inform simulations of the expected AO performance it is necessary to know the number of potential natural guide stars (NGS) at different limiting magnitudes within the EAGLE patrol field. In Table 3 we summarise the equatorial co-ordinates ($\alpha$, $\delta$; J2000.0) of a selection of likely target fields, together with their galactic latitudes ($l$) and the number of NGS (with $R$-band magnitudes brighter than 16, 17, and 18) that are within a 5' circular field. The NGS were drawn from searches of the United States Naval Observatory (USNO) A2.0 Catalogue [31].

**Table 3: Summary of available natural guide stars (NGS) for example E-ELT target fields**

| Target field | Science Topic | $\alpha$ [$^h$, $^m$, $^s$] | $\delta$ [°,',"] | $l$ [°] | <#NGS> $R < 16$ | $R < 17$ | $R < 18$ |
|---|---|---|---|---|---|---|---|
| XMM-LSS | Galaxy evolution | 00 24 40 | –04 30 00 | –66.5 | 2.5 | 3.7 | 5.8 |
| ELAIS S1 | Galaxy evolution | 00 34 44 | –43 28 01 | +73.3 | 2.5 | 4.8 | 23.4 |
| NGC 1097 | AGN | 02 46 19 | –30 16 29 | –64.7 | 1.6 | 3.2 | 20.5 |
| CDFS | Galaxy evolution | 03 32 28 | –27 48 30 | –54.4 | 2.8 | 4.8 | 13.4 |
| NGC 1365 | AGN | 03 33 36 | –36 08 28 | –54.6 | 2.5 | 4.2 | 18.4 |
| COSMOS | Galaxy evolution | 10 00 29 | +02 12 21 | +42.1 | 3.0 | 5.4 | 8.6 |
| Cen A | Stellar populations | 13 25 28 | –43 01 09 | +19.4 | 14.6 | 32.8 | 120.1 |
| M83 | Stellar populations | 13 37 01 | –29 51 59 | +32.0 | 7.0 | 14.0 | 58.3 |
| ELAIS N1 | Galaxy evolution | 16 10 01 | +54 30 36 | +4.1 | 3.7 | 6.3 | 10.0 |
| Abell 2218 | Galaxy evolution | 16 35 54 | +66 13 00 | +38.1 | 4.8 | 8.9 | 15.9 |
| NGC 7469 | AGN | 23 03 16 | +08 52 26 | –45.5 | 3.5 | 5.7 | 8.7 |

In the case of the spatially-extended, high-redshift survey fields (e.g. the two ELAIS fields), the USNO data were restricted to a 1 deg² area of the sky centred on the co-ordinates in Table 3. From these subsets, 100 randomly centred 5' fields were created and the number of NGS above a given brightness was found for each; the results in Table 3 represent the mean values from the 100 random fields. Figure 7 shows the number of NGS as a function of limiting $R$-band magnitude – for the majority of the example fields there are fewer than ten NGS in the 5' field-of-view with $R < 17$, ruling out a potential segmented-MCAO architecture. Of course, there are large variations in the density of potential NGS, depending on the galactic latitude of the target field.

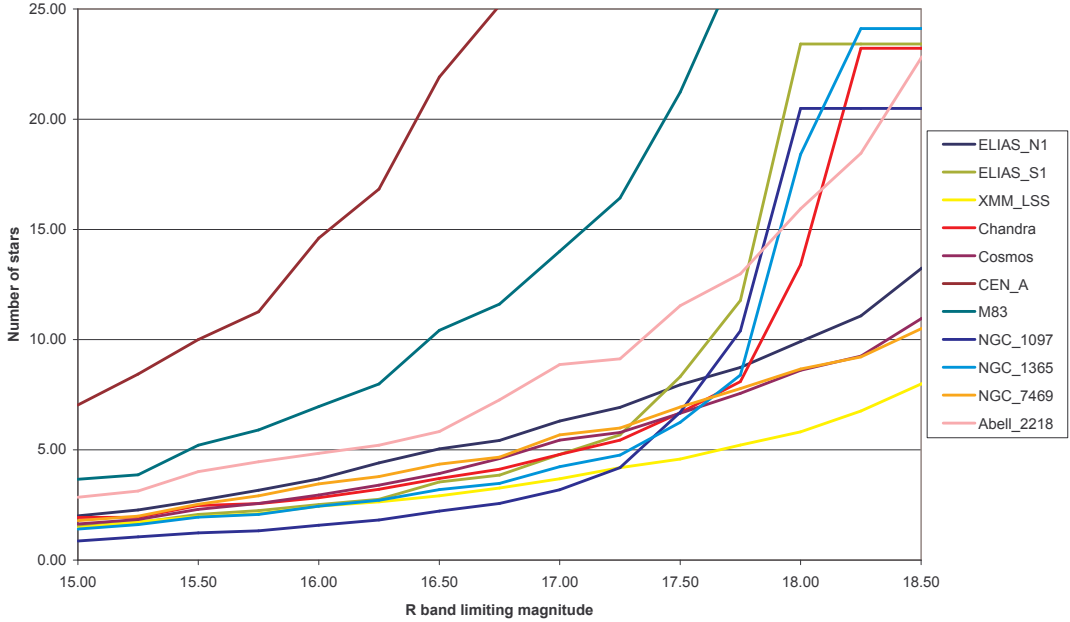

**Figure 7:** Number of natural guide stars (NGS) available in example 5' diameter EAGLE fields, as a function of $R$-band limiting magnitude.

## 8. SUMMARY

The baseline science requirements for the EAGLE Phase A study are listed in Table 4. The combination of the unprecedented primary aperture of the E-ELT with the large multiplex and excellent AO correction from EAGLE will yield huge gains in sensitivity and efficiency over existing facilities.

**Table 4: Baseline EAGLE science requirements**

| Parameter | Requirement |
|---|---|
| Patrol field | ≥ 5' diameter |
| Science (IFU) sub-field | ≥ 1.5" x 1.5" |
| Multiplex | ≥ 20 |
| Spatial resolution | ≥ 30% EE in 75 mas ($H$-band) |
| Spectral resolution | 4,000 & 10,000 |
| Wavelength coverage | 0.8 – 2.45 μm |
| Clustering/tiling | Distributed & clustered targets + the ability to map contiguous regions |

*Acknowledgements:* We thank Mathieu Puech for his simulation results in advance of publication and Miska LeLouarn for kindly providing his PSF simulations.

**REFERENCES**

1. Cuby, J.-G. et al., 2008, these proceedings
2. Förster Schreiber, N. et al., 2006, ApJ, 645, 1062
3. Nesvadba, N. et al., 2008, A&A, 479, 67
4. Reddy, N. et al., 2006, ApJ, 653, 1004
5. Reddy, N. et al., 2008, ApJS, 175, 48
6. Förster Schreiber, N. et al., 2004, ApJ, 616, 40
7. Marchesini, D. et al., 2007, ApJ, 656, 42
8. Puech, M. et al., 2008, MNRAS, in press
9. Puech , M. et al., 2008, these proceedings
10. Iye, M. et al., 2006, Nature, 443, 186
11. Le Delliou, M. et al., 2005, MNRAS, 357, L11
12. Le Delliou, M. et al., 2006, MNRAS, 365, 712
13. Belokurov, V. et al., 2006, ApJ, 642, L137
14. Tolstoy, E. et al., 2001, MNRAS, 327, 918
15. Koch, A. et al., 2007, AJ, 133, 270
16. Battaglia, G. et al., 2008, MNRAS, 383, 183
17. Munari, U. et al., 2001, BaltA, 10, 613
18. Steinmetz, M. et al., 2006, AJ, 132, 1645
19. Massey, P. et al., 1989, AJ, 98, 1305
20. Lorente, N. P. F. et al., 2006, SPIE, 6274, 44
21. Bower, R. G. et al., 2006, MNRAS, 370, 645
22. Croton, D. J. et al., 2006, MNRAS, 365, 11
23. Magorrian, J. et al., 1998, AJ, 115, 2285
24. Gebhardt, K. et al., 2000, ApJ, 539, L13
25. McLure, R. J. et al., 2006, MNRAS, 368, 1395
26. Peng, C. Y. et al., 2006, ApJ, 649, 616
27. Lauer, T. R. et al., 2007, ApJ, 670, 249
28. Alexander, D. M. et al., 2008, AJ, 135, 1968
29. Netzer, H. et al., 2006, A&A, 453, 525
30. Prieto, M. A. et al., 2005, AJ, 130, 1472
31. Monet, D. et al., *USNO-A2.0: A Catalog of Astrometric Standards*, http://www.nofs.navy.mil/projects/pmm/